\documentclass[prl,showpacs,aps,twocolumn]{revtex4}
\usepackage{graphicx}
\usepackage{epstopdf}
\usepackage{dcolumn}
\usepackage{bm}
\usepackage{color}
\usepackage{ulem}
\usepackage{amsfonts,amsthm}
\usepackage{amsmath}
\usepackage{amssymb}
\begin{document}
\newcommand{\bR}{\mbox{\boldmath $R$}}
\newcommand{\tr}[1]{\textcolor{red}{#1}}
\newcommand{\trs}[1]{\textcolor{red}{\sout{#1}}}
\newcommand{\tb}[1]{\textcolor{blue}{#1}}
\newcommand{\tbs}[1]{\textcolor{blue}{\sout{#1}}}
\newcommand{\Ha}{\mathcal{H}}
\newcommand{\mh}{\mathsf{h}}
\newcommand{\mA}{\mathsf{A}}
\newcommand{\mB}{\mathsf{B}}
\newcommand{\mC}{\mathsf{C}}
\newcommand{\mS}{\mathsf{S}}
\newcommand{\mU}{\mathsf{U}}
\newcommand{\mX}{\mathsf{X}}
\newcommand{\sP}{\mathcal{P}}
\newcommand{\sL}{\mathcal{L}}
\newcommand{\sO}{\mathcal{O}}
\newcommand{\la}{\langle}
\newcommand{\ra}{\rangle}
\newcommand{\ga}{\alpha}
\newcommand{\gb}{\beta}
\newcommand{\gc}{\gamma}
\newcommand{\gs}{\sigma}
\newcommand{\vk}{{\bm{k}}}
\newcommand{\vq}{{\bm{q}}}
\newcommand{\vR}{{\bm{R}}}
\newcommand{\vQ}{{\bm{Q}}}
\newcommand{\vga}{{\bm{\alpha}}}
\newcommand{\vgc}{{\bm{\gamma}}}
\newcommand{\mb}[1]{\mathbf{#1}}
\arraycolsep=0.0em
\newcommand{\Ns}{N_{\text{s}}}
%

\title{
Charge Order in a Two-Dimensional Kondo Lattice Model
}

\author{
Takahiro Misawa, Junki Yoshitake, and Yukitoshi Motome
}

\affiliation{
Department of Applied Physics, University of Tokyo,
7-3-1 Hongo, Bunkyo-ku, Tokyo, 113-8656, Japan
}

\date{\today}

\begin{abstract}
The possibility of charge order is theoretically examined for the Kondo lattice model in two dimensions, which does not include bare repulsive interactions. 
Using two complementary numerical methods, we find that charge order appears at quarter filling in an intermediate Kondo coupling region. 
The charge ordered ground state is an insulator exhibiting an antiferromagnetic order at charge-poor sites, while the paramagnetic charge-ordered state at finite temperatures is metallic with pseudogap behavior.  
We confirm that the stability of charge order is closely related with the local Kondo-singlet formation at charge-rich sites. 
Our results settle the controversy on charge order in the Kondo lattice model in realistic spatial dimensions. 
\end{abstract}


\maketitle

Charge order (CO), a periodic spatial modulation of electron density which breaks the lattice translational symmetry, is a fundamental electronic state of matter. 
Since it was argued for the Verwey transition in magnetite~\cite{Verwey1939}, 
CO has been ubiquitously found in a broad range of correlated electron systems, such as organic conductors~\cite{Seo2004}, transition metal compounds~\cite{ImadaRMP}, and $f$-electron compounds~\cite{YbAs,YbIrSi}.  
One of the obvious origins of CO is the intersite Coulomb interaction, which prevents electrons from coming close to each other. 
For instance, a phase transition from paramagnetic (PM) metal to CO insulator takes place in an extended Hubbard model at quarter filling while increasing nearest-neighbor repulsive interaction{~\cite{Hubbard1978}}. 

An alternative, interesting possibility of CO was proposed for a Kondo lattice model which does not include any bare repulsive interaction between electrons [see Eq.~(\ref{eq:H})]~\cite{Hirsch}. 
The CO instability was deduced from a perturbation expansion in the limit of large onsite Kondo coupling $J$. 
The second-order perturbation in terms of the electron hopping $t$ gives rise to an effective repulsion between neighboring electrons $\propto t^2/J$, whose competition with the kinetic energy $\propto t$ was expected to cause CO at quarter filling in an intermediate $t/J$ regime. 
After the interesting proposal, the ground state of the Kondo lattice model in one dimension was studied numerically, but such CO instability was not found~\cite{Kondo_RMP,McCulloch,Shibata2011}. 
Recently, however, the problem was reexamined in infinite dimensions by employing the dynamical mean-field theory, and the instability toward CO was found near quarter filling~\cite{Otsuki,Peters}. 
Thus, the conclusions are contradicting between one and infinite dimensions, and it is unclear whether CO is realized in more realistic two and three dimensions. 
This is a fundamental issue related with the unique mechanism of CO in the Kondo lattice model, not seen in Hubbard-type models, and potentially relevant to understand CO phenomena in $f$-electron compounds. 

In this Letter, we explore CO in the quarter-filled Kondo lattice model on a square lattice by using two complementary methods: 
variational Monte Carlo (VMC) method~\cite{TaharaVMC_Full} for the ground state 
and cellular dynamical mean-field theory~(CDMFT)~\cite{KotliarCDMFT} for finite temperatures ($T$).
In the VMC simulation, the optimization of large number of variational parameters and careful system-size analysis allow us to obtain the ground state by seriously taking account of both spatial and quantum correlations. 
On the other hand, in CDMFT, we employ a continuous-time Monte Carlo method~\cite{GullRMP} for the impurity solver, which enables us to obtain finite-$T$ properties with fully including dynamical correlations as well as spatial correlations within the cluster. 
The results by the two methods provide convincing evidence for the appearance of CO. 
Furthermore, we show that the CO ground state is insulating and accompanied by an antiferromagnetic (AF) order.
At finite $T$, the charge gap closes as AF order disappears, while a pseudogap remains in the PM CO phase. 
We confirm that CO is stabilized by the local Kondo-singlet formation by studying the effect of an intersite AF exchange interaction.
Our results suggest that the Kondo lattice model accommodates CO in more than one dimension. 

\begin{figure}[htb!]
  \begin{center}
    \includegraphics[width=8.5cm,clip]{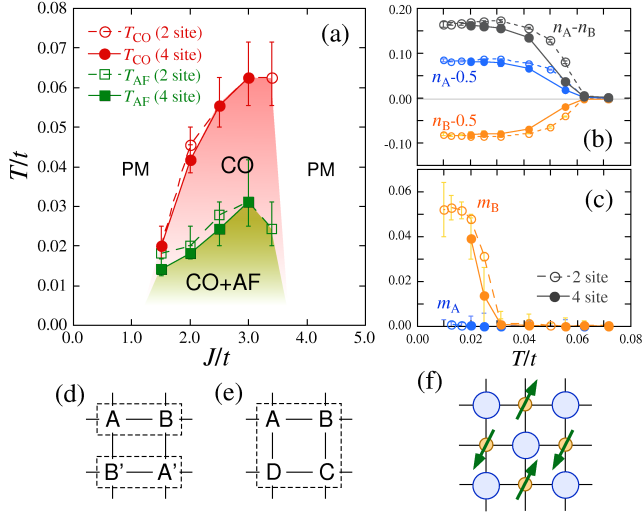}
  \end{center}
\caption{(color online). 
(a) Finite-$T$ phase diagram for the model (\ref{eq:H}) at quarter filling obtained by CDMFT. 
$T_{\mathrm{CO}}$ and $T_\mathrm{{AF}}$ denote the critical temperatures for charge ordering and antiferromagnetic ordering, respectively. 
(b) $T$ dependencies of the local charge density in the A and B sublattice, $n_\mathrm{A}$ and $n_\mathrm{B}$, at $J/t=3$. 
(c) $T$ dependencies of the magnitude of local magnetic moment of conduction electrons in the A and B sublattice, $m_\mathrm{A}$ and $m_\mathrm{B}$, at $J/t=3$. 
The lines connecting the data points are the guides for eyes. 
(d)(e) Sublattice for the two- and four-site cluster calculations. See the text for details.
(f) Schematic picture of the CO phase with AF order at the lowest $T$. 
The size of circles represent the magnitude of local density and the arrows show the magnetic moments. 
}
\label{fig:cDMFT}
\end{figure}

We consider the Kondo lattice model on a square lattice, whose Hamiltonian is given by
\begin{equation}
{\cal H} = -t\sum_{\langle i,j\rangle,\sigma}(c_{i\sigma}^{\dagger}c_{j\sigma}+\mathrm{h.c.})
+J\sum_{i}\boldsymbol{S}^{c}_{i}\cdot\boldsymbol{S}^{f}_{i},
\label{eq:H}
\end{equation}
where $c_{i\sigma}^{\dagger}$ ($c_{i\sigma}$) is a creation (annihilation) operator of a conduction electron with spin $\sigma$ at $i$th site.
The spin operator of conduction (localized) electron at $i$th site is defined by 
$\boldsymbol{S}^{c}_{i}={\frac12}\sum_{\sigma,\sigma^{\prime}}
c^{\dagger}_{i \sigma}\boldsymbol{\sigma}_{\sigma\sigma^{\prime}}c_{i \sigma^{\prime}}$ 
($\boldsymbol{S}^{f}_{i}={\frac12}\sum_{\sigma,\sigma^{\prime}}
f^{\dagger}_{i \sigma}\boldsymbol{\sigma}_{\sigma\sigma^{\prime}}f_{i \sigma^{\prime}}$; 
$f_{i\sigma}^\dagger$ and $f_{i\sigma}$ are creation and annihilation operators for the localized electrons, respectively), 
where $\boldsymbol{\sigma}_{\sigma\sigma^{\prime}}$ is a vector of the Pauli matrices.
The first term describes the hopping of conduction electrons between {the} nearest-neighbor sites $\langle i,j \rangle$ on the square lattice, and the second term represents the onsite AF Kondo coupling between conduction and localized spins ($J>0$). 
In the following, we set the lattice constant $a=1$ and focus on the quarter filling case, namely, at $n^c = \frac{1}{N_\mathrm{s}} \sum_{i \sigma} \langle c_{i \sigma}^\dagger c_{i \sigma} \rangle = 1/2$ ($N_\mathrm{s} = L \times L$ is the system size). 

We study the possibility of CO in the model (\ref{eq:H}) by two complementary methods, VMC and CDMFT. 
In the VMC calculations, we study the ground state properties by employing a generalized BCS wave function with the quantum number projection and the Gutzwiller and Jastrow factors; 
$|\psi\ra = \sP_{\rm G}\sP_{\rm J}\sL^{S}|\phi_{\rm pair}\ra$. 
Here, $\sL^{S}$ is the spin projection operator to the total spin $S$ subspace; 
$\sP_{\rm G}$ and $\sP_{\rm J}$ are the Gutzwiller and Jastrow factors, respectively~\cite{TaharaVMC_Full}.
The spin projection is performed onto the $S=0$ singlet subspace, except for the case explicitly mentioned. 
The Gutzwiller factor penalizes the double occupation of electrons by $\sP_{\text{G}} = \exp(-\sum_{i,\lambda} g_i^\lambda n_{i \uparrow}^\lambda n_{i \downarrow}^\lambda)$ where $n_{i\sigma}^\lambda = \lambda_{i \sigma}^\dagger \lambda_{i \sigma}$ ($\lambda=c, f$); 
we take $g_{i}^{f}=\infty$ for localized $f$ electrons.
The Jastrow factor is introduced only for conduction electrons as 
$\sP_{\text{J}} = \exp(-\frac{1}{2} \sum_{i,j} v_{ij}^{c} n_{i}^{c} n_{j}^{c})$,
where $n_{i}^{c} = \sum_\sigma n_{i \sigma}^c$. 
The one-body part $|\phi_{\rm pair}\ra$ is the generalized pairing wave function defined as
$|\phi_{\rm pair}\ra=(\sum_{\lambda,\nu=c,f}\sum_{i,j=1}^{\Ns}f_{ij}^{\lambda\nu} \lambda_{i \uparrow}^\dag \nu_{i \downarrow}^\dag)^{N_e/2} |0 \ra$, 
where $N_e$ is the number of electrons (including $f$ electrons). 
In this study, we restrict the variational parameters, $g_{i}^{c}$, $v_{ij}^{c}$, and $f_{ij}^{\lambda\nu}$, to have $2\times2$ sublattice structure. 
All the variational parameters are simultaneously optimized by using the stochastic reconfiguration method~\cite{TaharaVMC_Full,Sorella}.
Our variational wave function $|\psi\ra$ can flexibly describe CO, AF, and PM state on equal footing.
The calculations are done up to $12\times 12$ sites; 
to reduce the finite-size effects, we choose appropriate boundary conditions for each system size so as to satisfy the closed-shell condition in the noninteracting case $J=0$.
In addition to the ground state by VMC, we calculated finite-$T$ properties by CDMFT for two-site ($1\times 2$) and four-site ($2\times 2$) clusters [see Figs.~\ref{fig:cDMFT}(d) and \ref{fig:cDMFT}(e)]. 
For the two-site case, we effectively consider a $2\times 2$ sublattice order by configuring the Green's function with flipped spins so as to accomodate the ordering pattern found in the four-site cluster [Fig.~\ref{fig:cDMFT}(d)]. 
To solve the effective impurity problems for the clusters, we employ the continuous-time Monte Carlo technique~\cite{GullRMP}. 
Typically, we repeat the CDMFT loops for 50 times for convergence, and performed $10^6$-$10^7$ Monte Carlo measurements per loop.

Let us first discuss finite-$T$ properties. 
Figure~\ref{fig:cDMFT}(a) shows the finite-$T$ phase diagram obtained by the CDMFT calculations. 
The result shows that the model in Eq.~(\ref{eq:H}) exhibits CO at quarter filling in the intermediate coupling region for $1.5 \lesssim J/t \lesssim 3.5$. 
CO is a two-sublattice checkerboard type, as shown in Fig.~\ref{fig:cDMFT}(f).
$T$ dependence of the charge disproportionation at $J/t=3$ is shown in Fig.~\ref{fig:cDMFT}(b). 
The continuous development of the order parameter $n_\mathrm{A}-n_\mathrm{B}$ as lowering $T$ suggests that this CO transition is of second order.
As shown in Fig.~\ref{fig:cDMFT}(a), the critical temperature $T_{\rm CO}$ grows rapidly as increasing $J/t$, whereas it suddenly disappears for $J/t \gtrsim 3.5$ in the $T$ range calculated. 
The behavior of $T_{\rm CO}$ is very similar to that obtained in infinite dimensions (corresponding to a single-site cluster calculation)~\cite{Otsuki}. 
Interestingly, $T_{\rm CO}$ does not largely depend on the cluster sizes in the present calculations. 

In the CO phase, an AF order appears in the low $T$ region. 
In Fig.~\ref{fig:cDMFT}(c), we show $T$ dependences of the local magnetic moments of conduction electrons at the charge-rich site, $m_\mathrm{A}$, and charge-poor site, $m_\mathrm{B}$. 
At the charge-poor site, a spin polarization develops continuously as lowering $T$, while no magnetic moment is seen at the charge-rich site.
The AF order is a four-sublattice one, as shown in Fig.~\ref{fig:cDMFT}(f); 
collinear N{\'e}el type magnetic moments appear at the charge-poor sites, while the charge-rich sites remain nonmagnetic. 
Strictly speaking, the critical temperature $T_{\rm AF}$ should be zero, as the model has $SU(2)$ symmetry in two dimensions~\cite{Mermin1966}; 
the finite $T_{\rm AF}$ is an artifact of CDMFT. 
The result, however, reflects the development of AF correlations toward $T=0$. 
Hence, the CDMFT results suggest that the ground state is the CO state with AF long-range order for $1.5 \lesssim J/t \lesssim 3.5$.

\begin{figure}[t!]
  \begin{center}
    \includegraphics[width=8.5cm,clip]{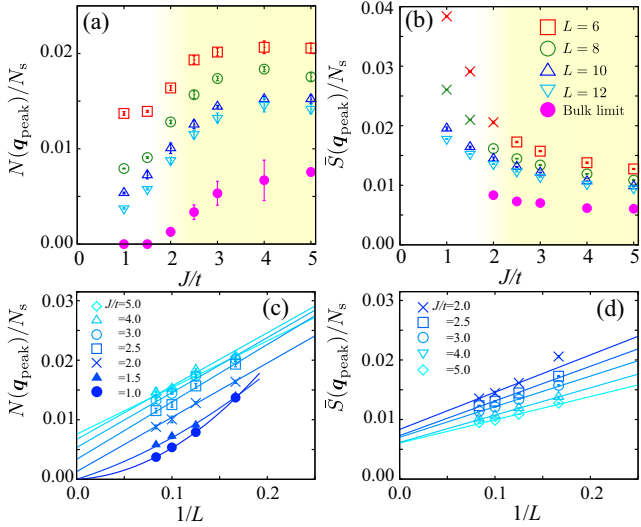}
  \end{center}
\caption{(color online). 
$J/t$ dependences of the peak values of (a) charge structure factor and (b) spin structure factor for system sizes $L=6,8,10$, and $12$ calculated by VMC.
System size extrapolations of the data are shown in (c) and (d). 
The extrapolated values in the bulk limit $L \to \infty$ are plotted in (a) and (b). 
}
\label{fig:VMCSqNq}
\end{figure}

Now, we investigate the ground state by the VMC calculations. 
Figures~\ref{fig:VMCSqNq}(a) and \ref{fig:VMCSqNq}(b) show $J/t$ dependences of the peaks of the charge structure factor and spin structure factor, which are defined as 
\begin{eqnarray}
N(\boldsymbol{q})&=&\frac{1}{N_{\rm s}}\sum_{i,j}
\langle (n^{c}_{i\uparrow}+n^{c}_{i\downarrow})
(n^{c}_{j\uparrow}+n^{c}_{j\downarrow}) \rangle 
e^{i\boldsymbol{q}\cdot(\boldsymbol{r}_{i}-\boldsymbol{r}_{j})},\\
S(\boldsymbol{q})&=&\frac{1}{3N_{\rm s}}\sum_{i,j}
\langle (\boldsymbol{S}^{c}_{i}+\boldsymbol{S}^{f}_{i})
\cdot(\boldsymbol{S}^{c}_{j}+\boldsymbol{S}^{f}_{j}) \rangle 
e^{i\boldsymbol{q}\cdot(\boldsymbol{r}_{i}-\boldsymbol{r}_{j})},
\end{eqnarray}
respectively. 
Here, $\boldsymbol{r}_i$ is the positional vector for the $i$th site. 
We find that $N(\boldsymbol{q})$ has a single peak at $\boldsymbol{q}_{\rm peak}=(\pi,\pi)$ for all the system sizes in the entire region of $J/t$. 
On the other hand, $S(\boldsymbol{q})$ shows two equivalent peaks at $\boldsymbol{q}_{\rm peak}=(\pi,0)$ and $(0,\pi)$ simultaneously, whereas it shows a single peak at $(\pi,0)$ or $(0,\pi)$ for small system 
sizes and for $J/t \lesssim 2$ [shown by crosses in Fig.~\ref{fig:VMCSqNq}(b)]; 
we plot $\bar{S}(\boldsymbol{q}_{\rm peak})=\{S(\pi,0)+S(0,\pi)\}/2$. 
The data well scale with $1/L$ as shown in Figs.~\ref{fig:VMCSqNq}(c) and \ref{fig:VMCSqNq}(d); 
the extrapolation to $L \to \infty$ indicates that CO accompanied by AF order appears in the bulk limit for $J/t\gtrsim 2$. 
The size extrapolation of $S(\boldsymbol{q})$ is not shown for $J/t<2$ because the magnetic ordering patterns depend on system sizes.
To clarify the magnetic order in this region, it is necessary to extend the sublattice for variational parameters; 
this is left for a future study. 
The VMC data indicate that the two-sublattice CO continuously appears for $J/t\gtrsim 1.5$, accompanied by the four-sublattice AF order, as illustrated in Fig.~\ref{fig:cDMFT}(f).

In the large $J/t$ region where $T_{\mathrm {CO}}$ disappears in CDMFT, we find that the VMC solution shows an instability toward a ferromagnetic state. 
Using the spin projection, we calculate the ground state energy as a function of the total spin $S$, $E(S)$. 
We found that $E(S)$ has local minima at $S=0$ and $S=N_\mathrm{s}/4$ for large $J/t$, 
which correspond to the singlet and ferromagnetic states, respectively. 
In the ferromagnetic state, conduction electrons form local singlets with localized spins at each site, 
and the rest unpaired localized moments are fully polarized. 
The result shows that $E(S=0) < E(S=N_\mathrm{s}/4)$ for $J/t \lesssim 4$, but the energy difference becomes smaller as increasing $J/t$;
$E(S=0) \sim E(S=N_\mathrm{s}/4)$ for $J/t \sim 5$~\cite{comment_size}. 
This implies that the system undergoes a transition from the CO+AF state to the ferromagnetic state at $J/t \sim 5$. 

Therefore, our complementary study by CDMFT and VMC clearly indicates that the CO+AF ground state appears in the intermediate $J/t$ region for $1.5 \lesssim J/t \lesssim 3.5$ in two dimensions. 
The result is in sharp contrast to the recent one obtained in infinite dimensions, in which the CO state {appears only at finite $T$ and} is replaced by a charge-uniform ferromagnetic state at low $T$ in the entire range of $J/t$~\cite{Peters}.

Our results suggest that CO is also stabilized in three dimensions because spatial fluctuations are suppressed as the spatial dimension becomes larger. 
It is, however, nontrivial what type of magnetic order appears in the ($\pi$,$\pi$,$\pi$)-CO state, as the magnetic moments at charge-poor sites suffer from geometrical frustration of the face-centered-cubic type. 
The interesting problem will be discussed elsewhere~\cite{Hayami}. 

\begin{figure}[t!]
  \begin{center}
    \includegraphics[width=8.5cm,clip]{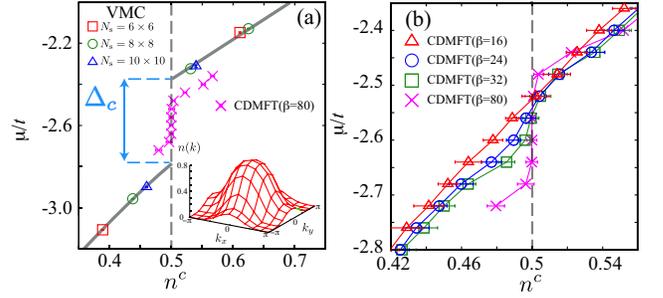}
  \end{center}
\caption{(color online). 
Chemical potential $\mu$ as a function of the electron density $n^c$ at $J/t=3$. 
Gray lines in (a) are the guides for eyes. 
The jump of $\mu$ at quarter filling $n^c=0.5$ corresponds to the charge gap $\Delta_{c}$ in the CO state.
CDMFT results are for the two-site cluster. 
The inset of (a) shows the momentum distribution function 
$n{({\boldsymbol{k}})}$ at $J/t=3$ {and $n^c=0.5$} obtained by VMC.
}
\label{fig:Ins}
\end{figure}

Next, to clarify the electronic structure of the obtained CO state, we calculate the chemical potential $\mu$ as a function of the electron density $n^c$.
In the VMC calculations, we calculate $\mu$ by using the relation 
$\mu(\bar{n}^c)=\{E(n_1^c)-E(n_2^c)\}/(n_1^c-n_2^c)$, 
where $E(n^c)$ is the total energy at $n^c$, and $\bar{n}^c=(n_1^c+n_2^c)/2$.
To reduce the finite-size effects, we choose $n_1^c$ and $n_2^c$ that satisfy the closed-shell conditions in the noninteracting case.
The results at $J/t=3$ are shown in Fig.~\ref{fig:Ins}(a). 
We find that $\mu$ shows a clear jump at quarter filling $n^c=0.5$. 
This indicates that the CO+AF ground state is an insulator. 
The charge gap is estimated by the jump, $\Delta_c \sim 0.4t $ at $J/t=3$. 
The insulating nature is also observed in the momentum distribution function, calculated by 
\begin{eqnarray}
n(\boldsymbol{k}) = \frac{1}{2N_\mathrm{s}} \sum_{i,j,\sigma} \langle c_{i \sigma}^\dagger c_{j \sigma} \rangle e^{i\boldsymbol{q}\cdot(\boldsymbol{r}_{i}-\boldsymbol{r}_{j})}. 
\end{eqnarray}
In the metallic state, $n(\boldsymbol{k})$ shows a discontinuity at the Fermi surface, while it does not show such singularity in the insulating state. 
The result for $J/t=3$ at quarter filling is shown in {the inset of} Fig.~\ref{fig:Ins}({a}). 
$n(\boldsymbol{k})$ shows a smooth $\boldsymbol{k}$ dependence, which supports that the system is insulating in the CO+AF ground state.

Correspondingly, in the CDMFT results, $\mu$ shows plateau-like behavior in the lowest-$T$ CO+AF phase, as shown in Fig.~\ref{fig:Ins}(a). 
Here, the CDMFT calculations are performed in the grand canonical ensemble, and hence, $n^c$ is calculated for a given $\mu$. 
As shown in Fig.~\ref{fig:Ins}(b), the plateau-like behavior is smeared out as raising $T$ in the intermediate-$T$ {PM CO} phase. 
The result suggests that the charge gap closes according to the disappearance of AF order. 
The PM CO phase at finite $T$ is metallic, while pseudogap behavior remains as kink-like behavior with an inflection point in the $\mu$-$n$ data shown in Fig.~\ref{fig:Ins}(b). 
The conclusion is apparently in contradiction with the recent result obtained in infinite dimensions, where an insulating PM CO state is predicted~\cite{Peters}.

\begin{figure}[tb!]
\begin{center}
\includegraphics[width=8.5cm,clip]{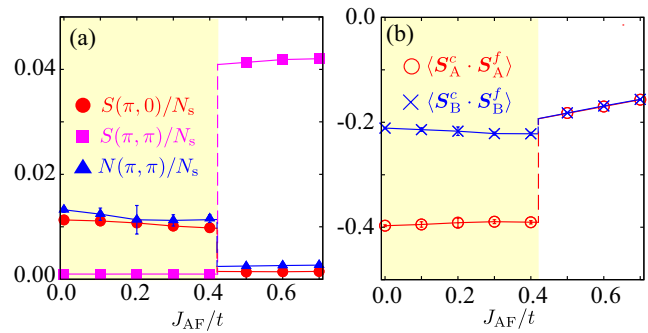}
\end{center}
\caption{(color online). 
(a)~$J_{\rm AF}$ dependences of the 
peak values of spin and charge structure factors at $J/t=3$. 
(b)~Local spin correlations at the charge-rich site (A) and charge-poor site (B). 
The data are obtained for $N_{\rm s}=12\times12$ by VMC.
}
\label{fig:SqNqS}
\end{figure}

Finally, let us examine the origin of CO. 
Originally, the possibility of CO in the Kondo lattice model was suggested by the second-order perturbation theory in terms of $t/J$~\cite{Hirsch}. 
In this argument, the local Kondo spin-singlet formed in the limit of $J/t \to \infty$ plays a key role in the stability of CO. 
This picture was also examined by an energetic argument considering the Kondo effect~\cite{Peters}. 
To confirm the importance of local spin-singlet formation, we numerically study the effect of a perturbation which destroys the local singlets. 
Here, we consider a perturbation to the model (\ref{eq:H}) by adding the AF exchange interaction between the localized spins, which is given by
\begin{eqnarray}
{\cal H}_{\mathrm{AF}} = J_{\mathrm{AF}} \sum_{\langle i,j \rangle} \boldsymbol{S}_i^f \cdot \boldsymbol{S}_j^f.
\end{eqnarray}
This term is expected to induce a N{\'e}el type AF order, which competes with the local spin-singlet formation. 
The VMC results while changing $J_{\mathrm{AF}}/t$ are shown in Fig.~\ref{fig:SqNqS}. 
As expected, the system exhibits a phase transition from the CO+AF state to N{\'e}el AF ordered state without CO at $J_{\mathrm{AF}}/t \simeq 0.42$. 
The transition point is estimated by the comparison of the ground state energy for two states. 
The transition is of first order; 
the peaks of $S(\boldsymbol{q})$ suddenly change from $(\pi,0)$ and $(0,\pi)$ to $(\pi,\pi)$, and simultaneously, the peak of $N(\boldsymbol{q})$ at $(\pi,\pi)$ is suppressed, as shown in Fig.~\ref{fig:SqNqS}(a). 
At the same time, as shown in Fig.~\ref{fig:SqNqS}(b), the local spin correlation $\langle \boldsymbol{S}_i^c \cdot \boldsymbol{S}_i^f \rangle$, which is a measure of the local singlet formation, is also suddenly suppressed. 
In particular, the local singlet at the charge-rich site is strongly suppressed at the transition. 
The results indicate that the local Kondo{-}singlet formation at the charge-rich sites plays an important role to stabilize CO, in consistent with the perturbation argument. 

To conclude, using two complementary numerical methods, VMC at $T=0$ and CDMFT at finite $T$, we have provided convincing evidences for CO in the Kondo lattice model in two dimensions. 
CO appears in the intermediate $J/t$ region, accompanied by a collinear AF order at the charge-poor sites (the charge-rich sites are nonmagnetic). 
We also found that the CO+AF state is insulating, whereas the PM CO state at finite $T$ is metallic while showing pseudogap behavior. 
We also confirmed that the local Kondo-singlet formation at the nonmagnetic sites plays an important role in stabilizing CO. 
Our results settle the controversy on the existence of CO in the Kondo {lattice} model in realistic dimensions. 

The present CO is of kinetic origin, and qualitatively different from the usual ones driven by bare intersite Coulomb repulsions. 
In the realistic compounds, however, the intersite Coulomb repulsions are also present, which may cooperatively work with the Kondo coupling on the tendency toward CO. 
Nevertheless, it will be possible to examine which mechanism is dominant by, e.g., applying magnetic field and pressure that affect the bandwidth, local Kondo coupling, and intersite interactions in a different way. 

The authors thank S.~Hoshino and H.~Kusunose for fruitful discussions. 
They also thank D.~Tahara and S.~Morita for providing us with efficient VMC codes. 
This research was supported by KAKENHI (No. 21340090, 23102708, and 24340076), Global COE Program ``the Physical Sciences Frontier", the Strategic Programs for Innovative Research (SPIRE), MEXT, and the Computational Materials Science Initiative (CMSI), Japan. 
Numerical calculation was partly carried out at the Supercomputer Center, Institute for Solid State Physics, Univ. of Tokyo.


\begin{thebibliography}{19}
\expandafter\ifx\csname natexlab\endcsname\relax\def\natexlab#1{#1}\fi
\expandafter\ifx\csname bibnamefont\endcsname\relax
  \def\bibnamefont#1{#1}\fi
\expandafter\ifx\csname bibfnamefont\endcsname\relax
  \def\bibfnamefont#1{#1}\fi
\expandafter\ifx\csname citenamefont\endcsname\relax
  \def\citenamefont#1{#1}\fi
\expandafter\ifx\csname url\endcsname\relax
  \def\url#1{\texttt{#1}}\fi
\expandafter\ifx\csname urlprefix\endcsname\relax\def\urlprefix{URL }\fi
\providecommand{\bibinfo}[2]{#2}
\providecommand{\eprint}[2][]{\url{#2}}

\bibitem[{\citenamefont{Verwey}(1939)}]{Verwey1939}
\bibinfo{author}{\bibfnamefont{E.~J.~W.} \bibnamefont{Verwey}},
  \bibinfo{journal}{Nature} \textbf{\bibinfo{volume}{144}},
  \bibinfo{pages}{327} (\bibinfo{year}{1939}).

\bibitem[{\citenamefont{Seo et~al.}(2004)\citenamefont{Seo, Hotta, and
  Fukuyama}}]{Seo2004}
\bibinfo{author}{\bibfnamefont{H.}~\bibnamefont{Seo}},
  \bibinfo{author}{\bibfnamefont{C.}~\bibnamefont{Hotta}}, \bibnamefont{and}
  \bibinfo{author}{\bibfnamefont{H.}~\bibnamefont{Fukuyama}},
  \bibinfo{journal}{Chem. Rev.} \textbf{\bibinfo{volume}{104}},
  \bibinfo{pages}{5005} (\bibinfo{year}{2004}).

\bibitem[{\citenamefont{Imada et~al.}(1998)\citenamefont{Imada, Fujimori, and
  Tokura}}]{ImadaRMP}
\bibinfo{author}{\bibfnamefont{M.}~\bibnamefont{Imada}},
  \bibinfo{author}{\bibfnamefont{A.}~\bibnamefont{Fujimori}}, \bibnamefont{and}
  \bibinfo{author}{\bibfnamefont{Y.}~\bibnamefont{Tokura}},
  \bibinfo{journal}{Rev. Mod. Phys.} \textbf{\bibinfo{volume}{70}},
  \bibinfo{pages}{1039} (\bibinfo{year}{1998}).

\bibitem[{\citenamefont{Ochiai et~al.}(1990)\citenamefont{Ochiai, Suzuki, and
  Kasuya}}]{YbAs}
\bibinfo{author}{\bibfnamefont{A.}~\bibnamefont{Ochiai}},
  \bibinfo{author}{\bibfnamefont{T.}~\bibnamefont{Suzuki}}, \bibnamefont{and}
  \bibinfo{author}{\bibfnamefont{T.}~\bibnamefont{Kasuya}},
  \bibinfo{journal}{J. Phys. Soc. Jpn.} \textbf{\bibinfo{volume}{59}},
  \bibinfo{pages}{4129} (\bibinfo{year}{1990}).

\bibitem[{\citenamefont{Hossain et~al.}(2005)\citenamefont{Hossain, Schmidt,
  Schnelle, Jeevan, Geibel, Ramakrishnan, Mydosh, and Grin}}]{YbIrSi}
\bibinfo{author}{\bibfnamefont{Z.}~\bibnamefont{Hossain}},
  \bibinfo{author}{\bibfnamefont{M.}~\bibnamefont{Schmidt}},
  \bibinfo{author}{\bibfnamefont{W.}~\bibnamefont{Schnelle}},
  \bibinfo{author}{\bibfnamefont{H.~S.} \bibnamefont{Jeevan}},
  \bibinfo{author}{\bibfnamefont{C.}~\bibnamefont{Geibel}},
  \bibinfo{author}{\bibfnamefont{S.}~\bibnamefont{Ramakrishnan}},
  \bibinfo{author}{\bibfnamefont{J.~A.} \bibnamefont{Mydosh}},
  \bibnamefont{and} \bibinfo{author}{\bibfnamefont{Y.}~\bibnamefont{Grin}},
  \bibinfo{journal}{Phys. Rev. B} \textbf{\bibinfo{volume}{71}},
  \bibinfo{pages}{060406} (\bibinfo{year}{2005}).

\bibitem[{\citenamefont{Hubbard}(1978)}]{Hubbard1978}
\bibinfo{author}{\bibfnamefont{J.}~\bibnamefont{Hubbard}},
  \bibinfo{journal}{Phys. Rev. B} \textbf{\bibinfo{volume}{17}},
  \bibinfo{pages}{494} (\bibinfo{year}{1978}).

\bibitem[{\citenamefont{Hirsch}(1984)}]{Hirsch}
\bibinfo{author}{\bibfnamefont{J.~E.} \bibnamefont{Hirsch}},
  \bibinfo{journal}{Phys. Rev. B} \textbf{\bibinfo{volume}{30}},
  \bibinfo{pages}{5383} (\bibinfo{year}{1984}).

\bibitem[{\citenamefont{Tsunetsugu et~al.}(1997)\citenamefont{Tsunetsugu,
  Sigrist, and Ueda}}]{Kondo_RMP}
\bibinfo{author}{\bibfnamefont{H.}~\bibnamefont{Tsunetsugu}},
  \bibinfo{author}{\bibfnamefont{M.}~\bibnamefont{Sigrist}}, \bibnamefont{and}
  \bibinfo{author}{\bibfnamefont{K.}~\bibnamefont{Ueda}},
  \bibinfo{journal}{Rev. Mod. Phys.} \textbf{\bibinfo{volume}{69}},
  \bibinfo{pages}{809} (\bibinfo{year}{1997}).

\bibitem[{\citenamefont{McCulloch et~al.}(2002)\citenamefont{McCulloch,
  Juozapavicius, Rosengren, and Gulacsi}}]{McCulloch}
\bibinfo{author}{\bibfnamefont{I.~P.} \bibnamefont{McCulloch}},
  \bibinfo{author}{\bibfnamefont{A.}~\bibnamefont{Juozapavicius}},
  \bibinfo{author}{\bibfnamefont{A.}~\bibnamefont{Rosengren}},
  \bibnamefont{and} \bibinfo{author}{\bibfnamefont{M.}~\bibnamefont{Gulacsi}},
  \bibinfo{journal}{Phys. Rev. B} \textbf{\bibinfo{volume}{65}},
  \bibinfo{pages}{052410} (\bibinfo{year}{2002}).

\bibitem[{\citenamefont{Shibata and Hotta}(2011)}]{Shibata2011}
\bibinfo{author}{\bibfnamefont{N.}~\bibnamefont{Shibata}} \bibnamefont{and}
  \bibinfo{author}{\bibfnamefont{C.}~\bibnamefont{Hotta}},
  \bibinfo{journal}{Phys. Rev. B} \textbf{\bibinfo{volume}{84}},
  \bibinfo{pages}{115116} (\bibinfo{year}{2011}).

\bibitem[{\citenamefont{Otsuki et~al.}(2009)\citenamefont{Otsuki, Kusunose, and
  Kuramoto}}]{Otsuki}
\bibinfo{author}{\bibfnamefont{J.}~\bibnamefont{Otsuki}},
  \bibinfo{author}{\bibfnamefont{H.}~\bibnamefont{Kusunose}}, \bibnamefont{and}
  \bibinfo{author}{\bibfnamefont{Y.}~\bibnamefont{Kuramoto}},
  \bibinfo{journal}{J. Phys. Soc. Jpn.} \textbf{\bibinfo{volume}{78}},
  \bibinfo{pages}{034719} (\bibinfo{year}{2009}).

\bibitem[{\citenamefont{Peters et~al.}(2013)\citenamefont{Peters, Hoshino,
  Kawakami, Otsuki, and Kuramoto}}]{Peters}
\bibinfo{author}{\bibfnamefont{R.}~\bibnamefont{Peters}},
  \bibinfo{author}{\bibfnamefont{S.}~\bibnamefont{Hoshino}},
  \bibinfo{author}{\bibfnamefont{N.}~\bibnamefont{Kawakami}},
  \bibinfo{author}{\bibfnamefont{J.}~\bibnamefont{Otsuki}}, \bibnamefont{and}
  \bibinfo{author}{\bibfnamefont{Y.}~\bibnamefont{Kuramoto}},
  \bibinfo{journal}{preprint (arXiv:1302.5467)}  (\bibinfo{year}{2013}).

\bibitem[{\citenamefont{Tahara and Imada}(2008)}]{TaharaVMC_Full}
\bibinfo{author}{\bibfnamefont{D.}~\bibnamefont{Tahara}} \bibnamefont{and}
  \bibinfo{author}{\bibfnamefont{M.}~\bibnamefont{Imada}}, \bibinfo{journal}{J.
  Phys. Soc. Jpn} \textbf{\bibinfo{volume}{77}}, \bibinfo{pages}{114701}
  (\bibinfo{year}{2008}).

\bibitem[{\citenamefont{Kotliar et~al.}(2001)\citenamefont{Kotliar, Savrasov,
  P\'alsson, and Biroli}}]{KotliarCDMFT}
\bibinfo{author}{\bibfnamefont{G.}~\bibnamefont{Kotliar}},
  \bibinfo{author}{\bibfnamefont{S.~Y.} \bibnamefont{Savrasov}},
  \bibinfo{author}{\bibfnamefont{G.}~\bibnamefont{P\'alsson}},
  \bibnamefont{and} \bibinfo{author}{\bibfnamefont{G.}~\bibnamefont{Biroli}},
  \bibinfo{journal}{Phys. Rev. Lett.} \textbf{\bibinfo{volume}{87}},
  \bibinfo{pages}{186401} (\bibinfo{year}{2001}).

\bibitem[{\citenamefont{Gull et~al.}(2011)\citenamefont{Gull, Millis,
  Lichtenstein, Rubtsov, Troyer, and Werner}}]{GullRMP}
\bibinfo{author}{\bibfnamefont{E.}~\bibnamefont{Gull}},
  \bibinfo{author}{\bibfnamefont{A.~J.} \bibnamefont{Millis}},
  \bibinfo{author}{\bibfnamefont{A.~I.} \bibnamefont{Lichtenstein}},
  \bibinfo{author}{\bibfnamefont{A.~N.} \bibnamefont{Rubtsov}},
  \bibinfo{author}{\bibfnamefont{M.}~\bibnamefont{Troyer}}, \bibnamefont{and}
  \bibinfo{author}{\bibfnamefont{P.}~\bibnamefont{Werner}},
  \bibinfo{journal}{Rev. Mod. Phys.} \textbf{\bibinfo{volume}{83}},
  \bibinfo{pages}{349} (\bibinfo{year}{2011}).

\bibitem[{\citenamefont{Sorella}(2001)}]{Sorella}
\bibinfo{author}{\bibfnamefont{S.}~\bibnamefont{Sorella}},
  \bibinfo{journal}{Phys. Rev. B} \textbf{\bibinfo{volume}{64}},
  \bibinfo{pages}{024512} (\bibinfo{year}{2001}).

\bibitem[{\citenamefont{Mermin and Wagner}(1966)}]{Mermin1966}
\bibinfo{author}{\bibfnamefont{N.~D.} \bibnamefont{Mermin}} \bibnamefont{and}
  \bibinfo{author}{\bibfnamefont{H.}~\bibnamefont{Wagner}},
  \bibinfo{journal}{Phys. Rev. Lett.} \textbf{\bibinfo{volume}{17}},
  \bibinfo{pages}{1133} (\bibinfo{year}{1966}).

\bibitem[{com()}]{comment_size}
\bibinfo{note}{We consider the system size with {$N_\mathrm{s} = 8\times 8$}
  under the periodic boundary condition in one direction and the antiperiodic
  boundary condition in the other direction, which satisfies the closed-shell
  condition in both the $S=0$ singlet state and the $S=N_\mathrm{s}/4$
  ferromagnetic state.}

\bibitem[{\citenamefont{Hayami et~al.}(2013)\citenamefont{Hayami, Misawa, and
  Motome}}]{Hayami}
\bibinfo{author}{\bibfnamefont{S.}~\bibnamefont{Hayami}},
  \bibinfo{author}{\bibfnamefont{T.}~\bibnamefont{Misawa}}, \bibnamefont{and}
  \bibinfo{author}{\bibfnamefont{Y.}~\bibnamefont{Motome}},
  \bibinfo{journal}{unpublished}  (\bibinfo{year}{2013}).

\end{thebibliography}

\noindent

\end{document}